# Structural, and Magnetic properties of flux free large FeTe Single crystal

P. K Maheshwari[1,2], Rajveer Jha[1], Bhasker Gahtori and V.P.S. Awana[1*]

[1]*CSIR-National Physical Laboratory, Dr. K. S. Krishnan Marg, New Delhi-110012, India*

[2]*AcSIR-National Physical Laboratory, New Delhi-110012, India*

ABSTRACT

We report synthesis of non superconducting parent compound of iron chalcogenide, i.e., FeTe single crystal by self flux method. The FeTe single crystal is crystallized in tetragonal structure with the P4/nmm space group. The detailed SEM (scanning electron microscopy) results showed that the crystals are formed in slab like morphology and are near (slight deficiency of Te) stoichiometric with homogenous distribution of Fe and Te. The coupled structural and magnetic phase transition is seen at around 70K in both electrical resistivity and magnetization measurements, which is hysteric ($\Delta T = 5K$) in nature with cooling and warming cycles. Magnetic susceptibility ($\chi$-T) measurements showed the magnetic transition to be of antiferromagnetic nature, which is substantiated by isothermal magnetization (M-H) plots as well. The temperature dependent electrical resistivity measured in 10kOe field in both in plane and out of plane field directions showed that the hysteric width nearly becomes double to $\Delta T =$ 10K, and is maximum for the out of plane field direction for the studied FeTe single crystal. We also obtained a sharp spike like peak in heat capacity Cp(T) measurement due to the coupled structural and magnetic order phase transitions.

***Corresponding Author**
Dr. V. P. S. Awana,
Principal Scientist
E-mail: awana@mail.npindia.org
Ph. +91-11-45609357, Fax-+91-11-45609310
Homepage awanavps.wenbs.com

INTRODUCTION

The discovery of Iron based superconductors had surely attracted great attention of the condensed matter physics community. These Iron based superconductors differ from phonon-mediated conventional superconductors in a way that their superconducting transition temperature ($T_c$) increases from 26K to as high as 55K, when nonmagnetic La in LaFeAs(O/F) is replaced by magnetic lanthanides, [1–4]. This somehow suggests to break the myth of contemplate that magnetic ions break Cooper pairs of superconductors [5]. The parent undoped phase of Fe based pnictide superconductors i.e., REFeAsO is antiferromagnetic (AFM), which arises from the nesting of spin density wave (SDW) type order [6]. The superconductivity has been achieved through suppressing long range magnetic order by charge carrier doping or external pressure [7]. The parent non superconducting phase Fe of chalcogenide superconductors, i.e., FeTe has also anti-ferromagnetic ordering as similar to the compounds of Fe pnictides [8]. The parent compound FeTe exhibits drastic drop at T~70 K in electrical resistivity and magnetic susceptibility with decreasing temperature, which is related to anti-ferromagnetic ordering and coupled first-order structural phase transition [9,10]. Clearly, the understanding of the ground state of both Fe pnictides and chalcogenide holds the key to shed light on mysterious superconductivity of these systems. Large single crystals of these compounds could be very useful in this regards.

Recently we reported on the growth of $FeSe_{1/2}Te_{1/2}$ single crystals without flux [11], which encouraged us to synthesize the FeTe single crystals through the same method. The iron chalcogenide single crystal materials were recently grown without flux by travelling floating zone technique and flux free growth [12]. The FeTe crystals haven grown earlier as well by self flux method but with help of high end furnaces employing travelling floating zone technique [13,14]. In this short article, we report the successful synthesis of FeTe single crystals without flux and that also in a normal tube furnace without any complicated heating schedules related to travelling-solvent floating zone technique. The synthesized FeTe single crystal is crystallized in the tetragonal structure with the space group P4/nmm and the characteristic first order phase transition at nearly T=70K has been observed in magnetic, electrical resistivity and heat capacity measurement. We believe such flux free large crystals if exposed to neutron scattering measurements will provide fruitful and important information about the nature of magnetic ordering in these compounds. Once the ground state is understood fully, one could map the transformation from magnetic (FeTe) to superconducting $FeSe_{1/2}Te_{1/2}$.

## EXPERIMENTAL DETAILS

The studied single crystals of FeTe were grown by a self flux melt growth method. We take high purity (99.99%) Fe, and Te powder in stoichiometric ratio and grind thoroughly in the argon atmosphere. Then mixed powder pelletized by applying uniaxial stress of 100 $kg/cm^2$ and sealed in an evacuated ($<10^{-3}$ Torr) quartz tube. The quartz capsules contained FeTe pallets were kept in the furnace for heating at temperature 450◦C with a rate of 2◦C/min for 4h and then rise temperature up to $1000^oC$ with a rate of 2◦C/min for 24h finally the furnace cooled slowly $10^0C/h$ down to room temperature. The obtained crystals have a platelet like shape with typical dimensions (2–1) cm × (1.0) cm. X-ray diifraction patterns are taken on Rigaku x-ray diffractometer using $CuK_\alpha$ line of 1.54184Å. The morphology and compositional analysis of the obtained single crystal is seen by scanning electron microscopy (SEM) images on a ZEISS-EVO MA-10 scanning electron microscope having coupled Energy Dispersive X-ray spectroscopy (EDAX). Electrical, magnetic and thermal (heat capacity) measurements were carried out on Quantum Design (QD) Physical Property Measurement System (PPMS) - 140kOe down to 2K.

## RESULTS AND DISCUSSION:

Figure 1 shows the Reitveld fitted and observed X-ray diffraction of gently crushed FeTe single crystal powder at room temperature. The compound is crystallized in the tetragonal structure with the space group P4/nmm. The Reitveld refined parameters are a= b=3.82(2)Å and c= 6.29(1)Å, which are in agreement with earlier repots [13, 14]. The atomic positions are allocated as Fe at coordinate position (0.75, 0.25, 0) and Te at (0.25, 0.25, 0.28) in FeTe compound. The Reitveld refined results clearly show that un-reacted excess Fe is not present in our FeTe crystals. The inset of the figure 1 shows the XRD pattern of palette like piece of FeTe single crystal. The observed peak at 2θ= 47.8 with indexed plane (200), demonstrates that the crystals are grown along *a*- plane direction. To verify the morphology and elemental analysis of the single phase FeTe single crystal, we performed the Scanning electron microscope (SEM) and EDX. The SEM images are shown in the figures 2(a-d); well shaped single crystals can be seen at the top of the surface of the sample. One can clearly see the planer structural growth of FeTe single crystals. EDX data has been taken for the selected area of FeTe single crystal, as shown in the figure 2(e). The elemental analysis showed Fe:Te =1:0.8 [see figure 2(f)], exhibiting slight deficiency and Te.  Selected area EDX mapping images are shown in figures 2 (g, h). The EDX mapping data for Fe and Te along with the combined superimpose confirm the uniform distribution of both elements across the sample.

Figure 3 shows the temperature dependence of the dc magnetic susceptibility for FeTe single crystal sample under 1kOe field in both cooling and warming process. The FeTe single crystal compound exhibits relatively high magnetic moment. Below 300K, the susceptibility continuously increases on decreasing temperature, and then shows a pronounced downturn at 70K, exhibits a clear anomaly at this temperature.  The hysteresis has been observed in cooling and warming process. During the warming cycle, the transition shifts towards the high temperature side. Inset of the figure 3 shows the isothermal magnetization curve at various temperatures 65, 80 and 200K. The magnetization curves are linearly dependent on magnetic field for all temperatures and no signal of ferromagnetic nature in synthesized FeTe single crystal is seen. This indicates the absence of un-reacted iron in our FeTe single crystal. Likely, it has been already stated as the FeTe is the in-plane AFM configuration, while in consecutive layers the Fe magnetic moments alternate in their directions. The magnetic moments of Fe atoms depend on the spin structure and the ferromagnetic Fe chains due to alternate length of the nearest Fe-Fe bonds. Furthermore, the magneto elastic terms are non-negligible the effective magnetic free energy generated by the lattice favors a first order transition [15]. Under pressure study shows the AFM unit cell deform to the orthorhombic crystal structure of space group Cmma, while the AFM alter to the monoclinic structure of P21/m space group [16]. This report suggests that the structural phase transition occurs with applied pressure in resulting lattice parameters are little bit short along spin-parallel alignment. Figure 4 represents the temperature dependence of the ac magnetic susceptibility for FeTe single crystal sample in at different ac field amplitudes. The ac susceptibility initially an increase with decreasing temperature sudden drop in the ac susceptibility moment has been observed at 70K and a clear anomaly at this temperature. This transition is due to coupled structural and magnetic phase transitions and the height increases with increasing amplitudes of ac field.

Figure 5a shows the temperature dependence of electrical resistance ρ(T) for FeTe single crystal in two protocols cooling and warming with and without magnetic field from 300K to down to 2K. The 10kOe magnetic field was applied in two directions in plane and out of plane of the FeTe single crystalline sample. Initially the electrical resistivity increases with decreasing temperature a sharp transition has been observed nearly at 70K latter on the resistivity decreases with decreasing temperature. The figure 5b is enlarging view of the same in the temperature range 78-55K. The structural phase transition near to 66K has been observed at cooling time this increases from 66K to 71K at the warming of FeTe single crystal sample. The transition width is nearly 5K, which is increasing with applied magnetic field from 0-10kOe as well as for the change in the direction of applied field. When the sample position rotated from 0-90 degree i.e. field parallel to the sample the transition width increases from 5K to 11K, i.e. clear indication of anisotropy in the temperature dependence electrical resistivity. Earlier studies were focused on the temperature dependence of resistivity anisotropy in FeTe single crystals twinned with uniaxial strain [17]. It has been already shown that the resistivity along AFM direction is larger than that along the Ferromagnetic direction, which is exhibits the opposite behavior to the 122 series of Iron Pnictides and chalcogenide compounds [17-19]. The large change in the resistivity anisotropy behavior of FeTe is probably caused by the large localized moment and the spin of the itinerant electrons.

The temperature dependent heat-capacity measurement from 300K down to 2K is shown in figure 6 for FeTe single crystal sample. The most interesting feature is the spike like transition encountered in FeTe at about 72K. The anomaly at around 72K in $C_P(T)$ is generally due to the reported structural phase transition and possible contribution of any magnetic state such as SDW. However, heat capacity transition due to the heat absorption and liberation from the first-order phase transition become apparent. This transition gives a change in entropy through local Fe moment in FeTe single crystal, the change in the entropy across the transition suggests that the major contribution to the entropy change at the phase transition.

In summery we synthesized FeTe single crystal through flux free technique the as synthesized FeTe single crystal is crystallized in the tetragonal structure with the P4/nmm space group. The structural phase transition at 70K has been confirmed from dc & ac magnetic susceptibility measurements. We observed that the transition width increases with filed as well as direction of applied magnetic field at the time cooling and warming. We also obtained spike like peak in heat capacity Cp(T) measurement due to the structural and magnetic phase transition.

ACKNOWLEDGEMENT:

Authors would like to thank their Director NPL-CSIR India for his intense interest in the present work. This work is financially supported by *DAE-SRC* outstanding investigator award scheme on search for new superconductors. P. K. Maheshwari thanks CSIR, India for research fellowship and AcSIR-NPL for Ph.D. registration.

REFERENCES:

1. Y. Kamihara, T. Watanabe, M. Hirano, and H. Hosono, J. Am. Chem. Soc. **130**, 3296 (2008).
2. Z.-A. Ren, W. Lu, J. Yang, W. Yi, X.-L. Shen, Z.-C. Li, G.-C. Che, X.-L. Dong, L.-L. Sun, F. Zhou, and Z.-X. Zhao, Chin. Phys. Lett. **25**, 2215 (2008).
3. X. H. Chen, T. Wu, G. Wu, R. H. Liu, H. Chen and D. F. Fang, Nature **453**, 761 (2008).
4. G. F. Chen, Z. Li, D. Wu, G. Li, W. Z. Hu, J. Dong, P. Zheng, J. L. Luo, and N. L. Wang, Phys. Rev. Lett. **100**, 247002 (2008).
5. A. A. Abrikosov and L. P. Gorkov, Zh. Eksp. Teor. Fiz. **39**, 781 (1960).
6. F. Ma and Z.Y. Lu, Phys. Rev. B **78**, 033111 (2008).
7. G. Garbarino, A. Sow, P. Lejay, A. Sulpice, P. Toulemonde, M. Mezouar and M. Nunez-Regueiro, Europhys. Lett. **86**, 27001 (2009).
8. M. Rotter, M. Tegel, and D. Johrendt, Phys. Rev. Lett. **101**, 107006 (2008).
9. M. Bendele, P. Babkevich, S. Katrych, S. N. Gvasaliya, E. Pomjakushina, K. Conder, B. Roessli, A. T. Boothroyd, R. Khasanov and H. Keller, Phys. Rev. B **82**, 212504 (2010).
10. L. Zhang, D. J. Singh and M. H. Du, Phys. Rev. B **79**, 012506 (2009).
11. R. Jha, P.K. Maheshwari, B. Gahtori, V.P.S. Awana, arXiv:1502.01813 (2015).
12. E. Pomjakushina, Sup. Sci. & Tech. **27**, 120501 (2014)
13. J. Wen, G. Xu, G. Gu, J.M. Tranquada, and R.J. Birgeneau, Rep. Prog. Phys. 74, 124503 (2011).
14. D. P. Chen and C.T. Lin, Sup. Sci. & Tech. 27, 103002 (2014).
15. W. Bao, Y. Qiu, Q. Huang, M. A. Green, P. Zajdel, M. R. Fitzsimmons, M. Zhernenkov, S. Chang, M. Fang, B. Qian, E. K. Vehstedt, Jinhu Yang, H. M. Pham, L. Spinu, and Z. Q. Mao, Phy. Rev. Let. **102**, 247001 (2009).

16. A. V. Fedorchenko, G. E. Grechnev, V. A. Desnenko, A. S. Panfilov, S. L. Gnatchenko, V. Tsurkan, J. Deisenhofer, A. Loidl, O. S. Volkova and A. N. Vasiliev, J. Phys.: Condens. Matter **23,** 325701(2011).
17. S. Ishida, M. Nakajima, T. Liang, K. Kihou, C. H. Lee, A. Iyo, H. Eisaki, T. Kakeshita, Y. Tomioka, T. Ito, and S. Uchida, Phys. Rev. Lett. **110**, 207001 (2013).
18. C. H. Dong, H. D. Wang, Z. J. Li, J. Chen, H. Q. Yuan, and M. H. Fang, Phys. Rev. B **84**, 224506 (2011).
19. M. D. Johannes and I. I. Mazin, Phys. Rev. B **79**, 220510(R) (2009).

**Figure Captions**

**Figure 1:** The room temperature observed and Reitveld fitted XRD patterns of crushed powder of FeTe single crystal. Inset is room temperature XRD pattern of FeTe single crystal.

**Figure 2:** (a-d) Photograph of FeTe single crystals (e-i) The EDX quantitative analysis graph of the FeTe single crystal.

**Figure 3:** DC magnetization plots for FeTe single crystal measured in the applied magnetic field, H= 1kOe. Inset shows Isothermal MH curve at 65K, 80K & 200K of FeTe single crystal.

**Figure 4:** The AC magnetic susceptibility in real (M’) situations at fixed frequency of 333Hz in varying amplitudes of 5–15Oe for FeTe single crystal.

**Figure 5:** (a) The temperature dependent electrical resistivity in temperature range 300-5K of FeTe single crystal in cooling and warming process under applied magnetic field 1Tesla in plane and out of plane. (b) Zoomed view of the same ρ(T) curve in temperature range 78-55K.

**Figure 6:** The temperature dependent heat capacity $C_p(T)$ in temperature range 250-5K of FeTe single crystal.

Figure 1:

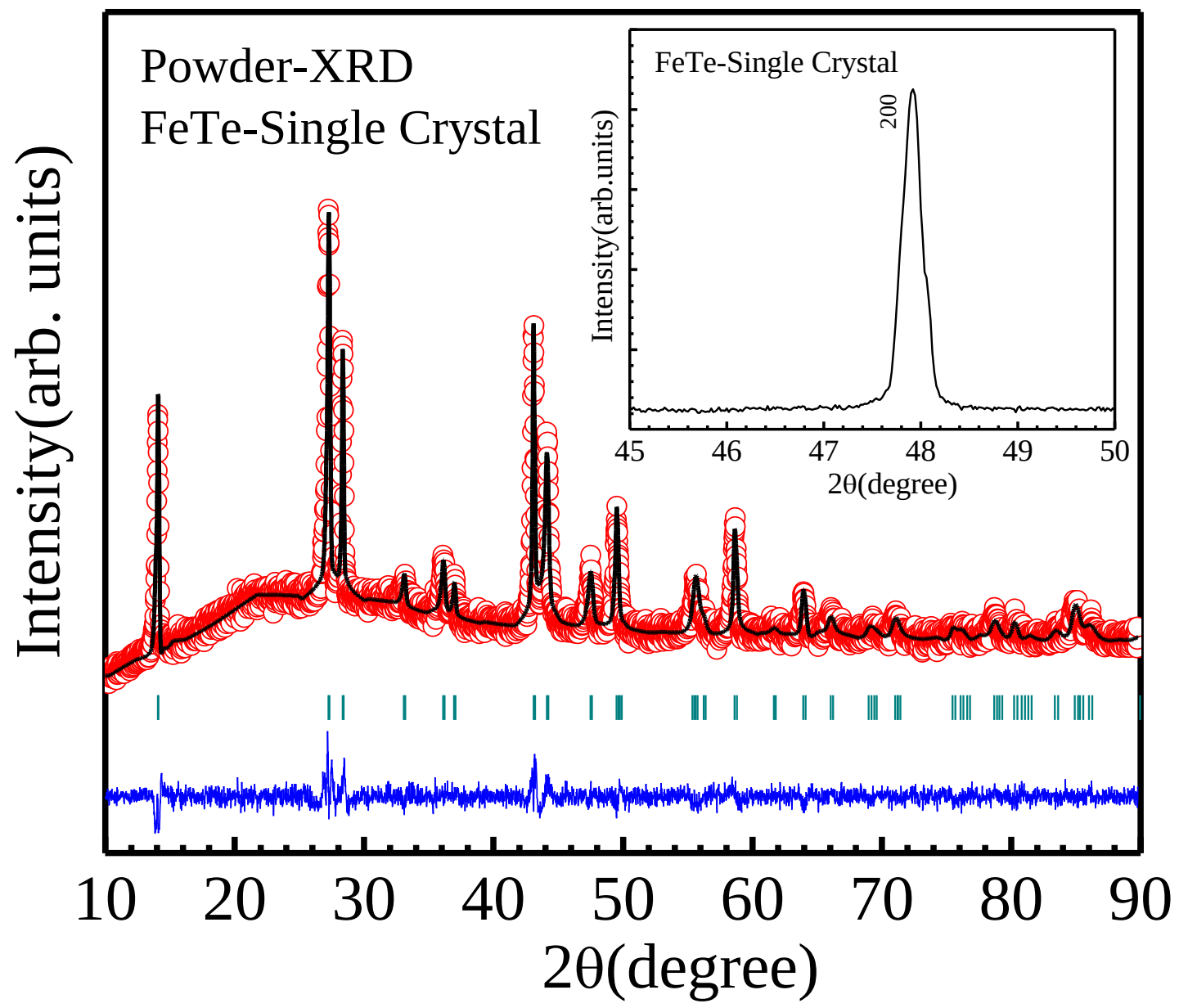

Figure 2:

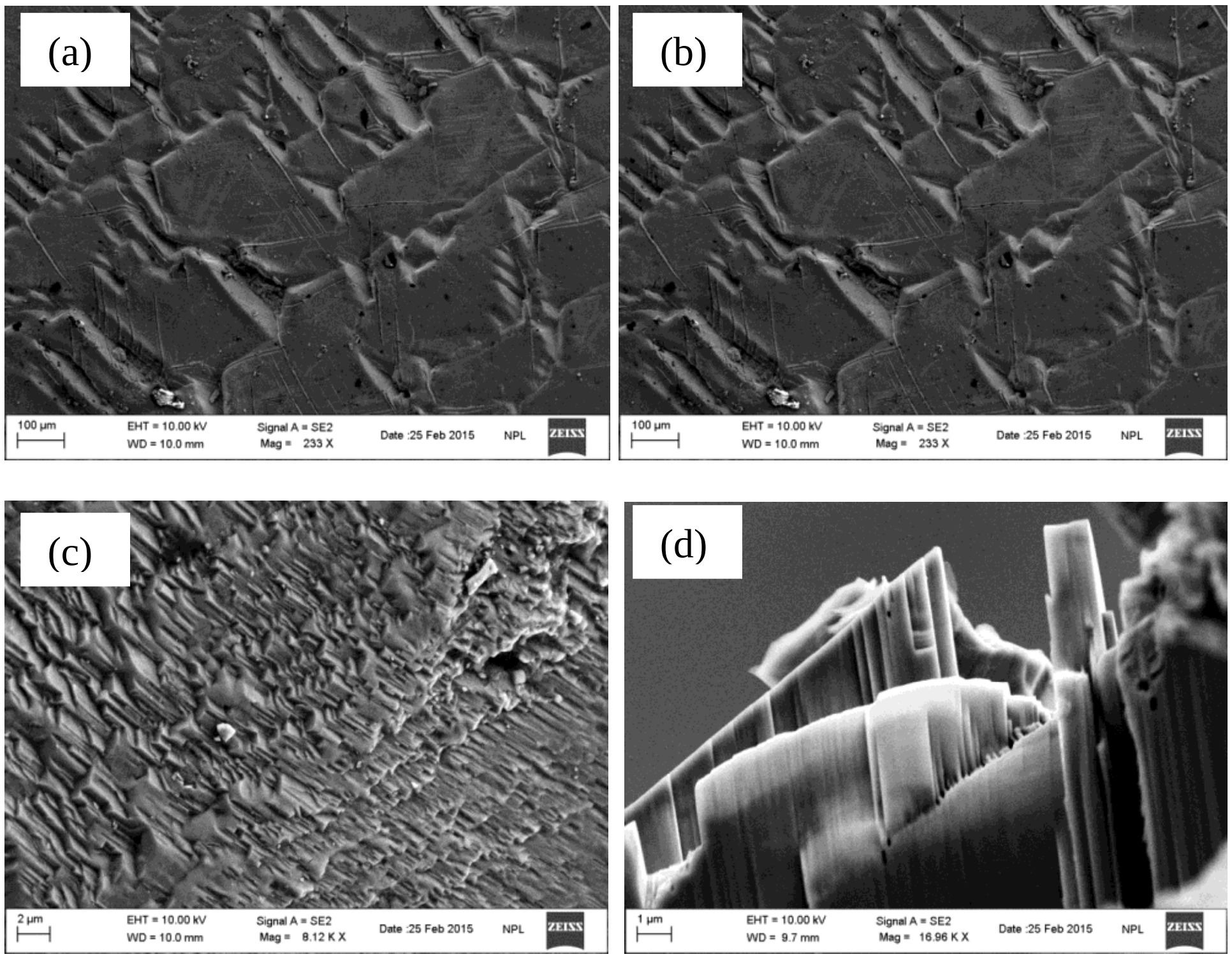

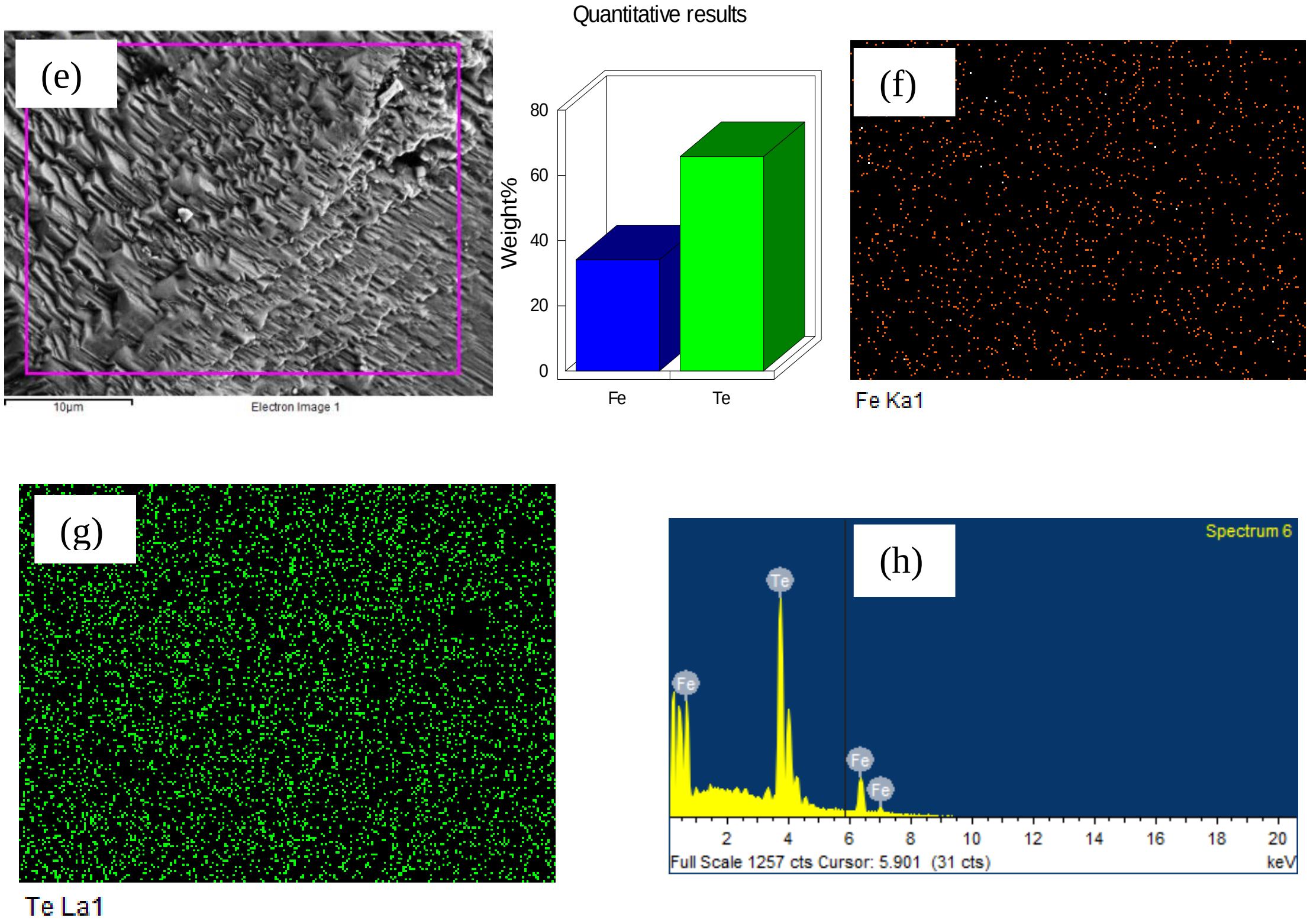
(e)
10μm
Electron Image 1
Quantitative results
Weight%
80
60
40
20
0
Fe
Te
(f)
Fe Ka1
(g)
Te La1
(h)
Spectrum 6
Te
Fe
Fe
Fe
Fe
2 4 6 8 10 12 14 16 18 20
Full Scale 1257 cts Cursor: 5.901 (31 cts)
keV

Figure 3:

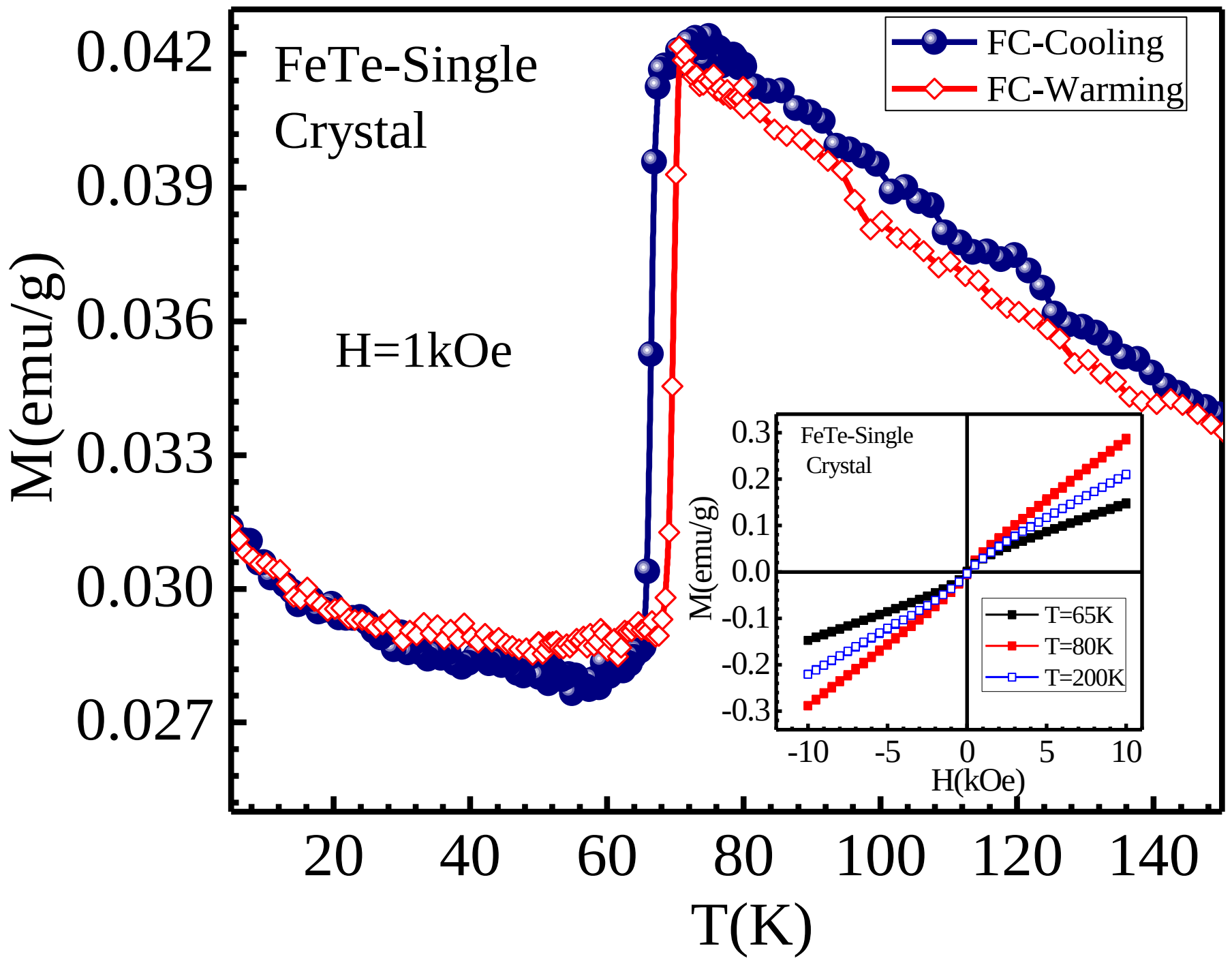

Figure 4:

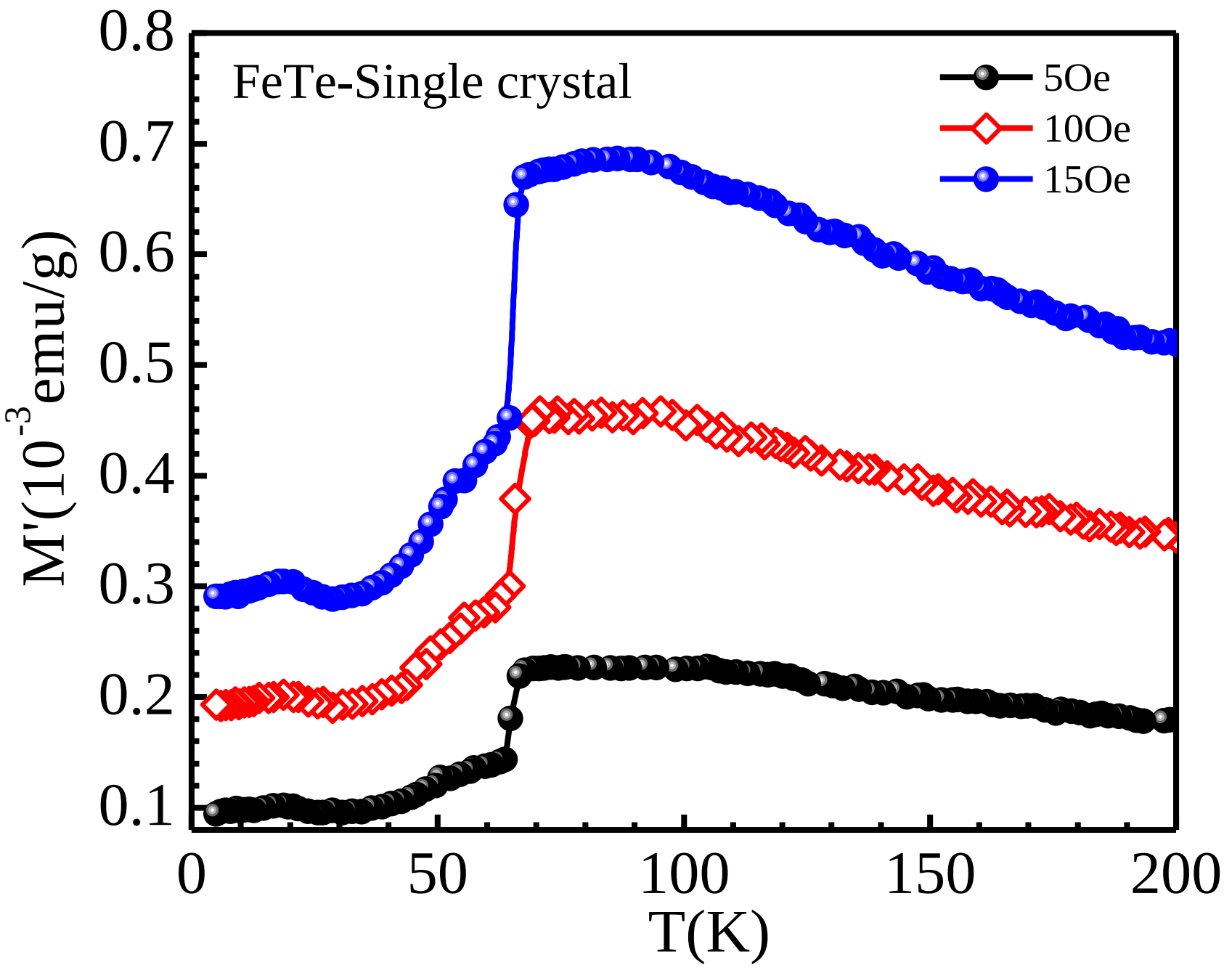

Figure 5:

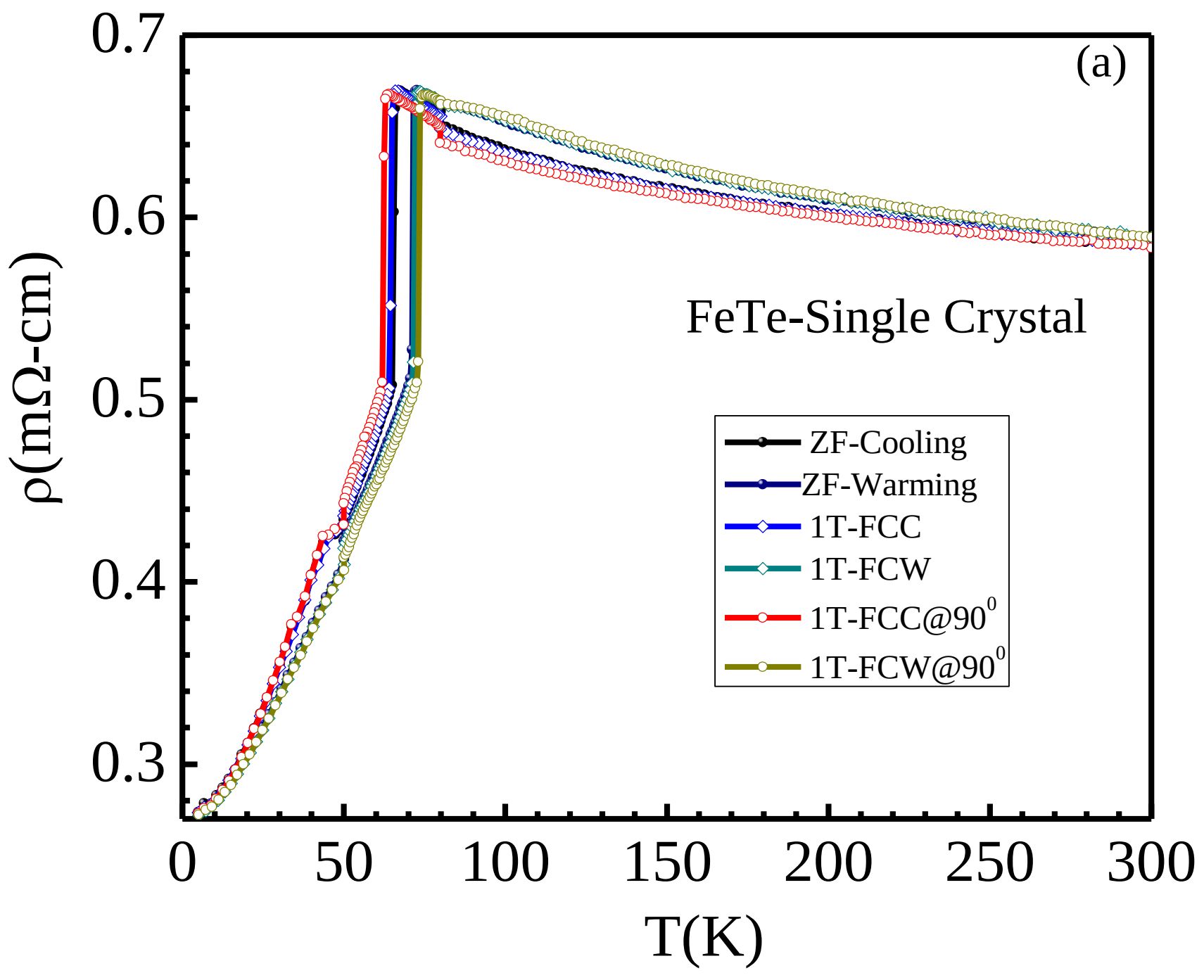

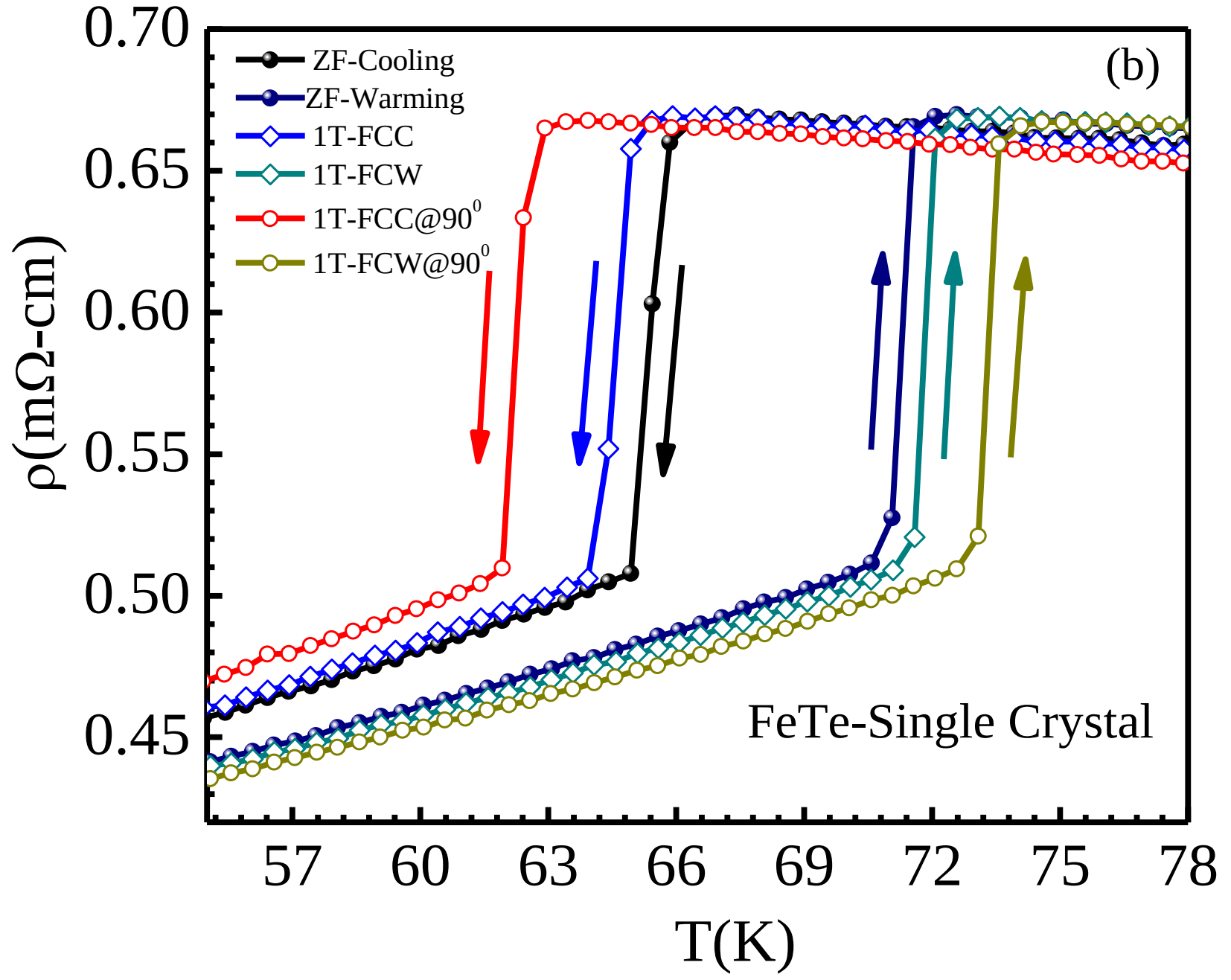

Figure 6:

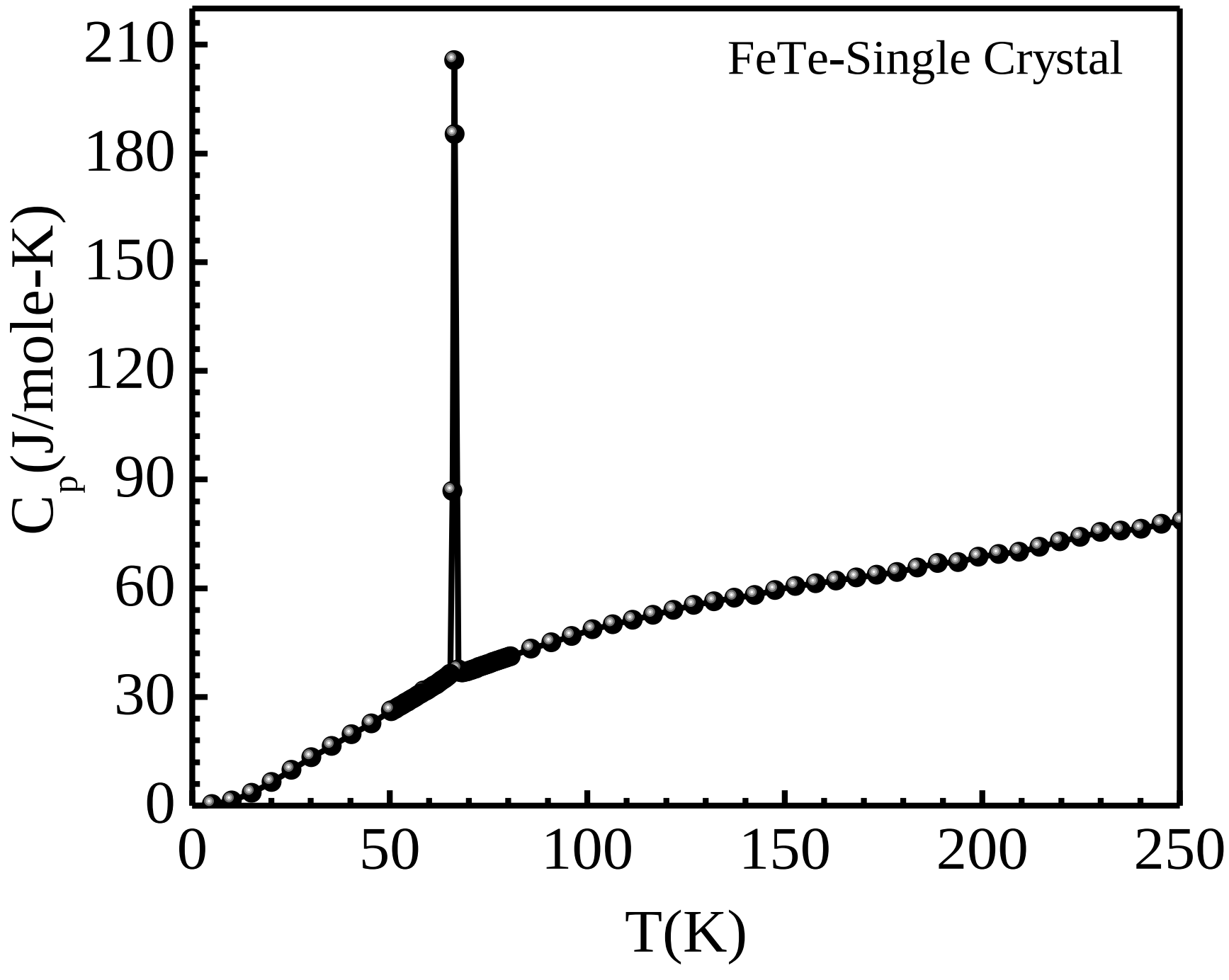